\documentclass[preprint,10pt]{elsarticle}

\usepackage{amssymb}

\usepackage{amsmath,array}
\usepackage{float}
\usepackage{hyperref}
\hypersetup{colorlinks,linkcolor=cyan,citecolor=cyan,urlcolor =[rgb]{0.0,0.0,0.5}}

\usepackage{geometry}
\journal{Physics Letters B 835 (2022) 137536, 
\href{https://doi.org/10.1016/j.physletb.2022.137536}{doi:10.1016/j.physletb.2022.137536}}

\begin{document}

\begin{frontmatter}



\title{Joint photon-electron Lorentz violation parameter plane from LHAASO data}


\author[1]{Ping He}
\author[1,2,3]{Bo-Qiang Ma}\ead{mabq@pku.edu.cn} 

\affiliation[1]{organization={School of Physics, Peking University},
            city={Beijing},
            postcode={100871}, 
            country={China}}
\affiliation[2]{organization={Center for High Energy Physics, Peking University},
            city={Beijing},
            postcode={100871}, 
            country={China}}
            
\affiliation[3]{organization={Collaborative Innovation Center of Quantum Matter},
            city={Beijing},
            country={China}} 

\begin{abstract}
The Large High Altitude Air Shower Observatory~(LHAASO) is one of the most sensitive gamma-ray detector arrays, whose ultrahigh-energy~(UHE) work bands not only help to study the origin and acceleration mechanism of UHE cosmic rays, but also provide the opportunity to test fundamental physics concepts such as Lorentz symmetry. LHAASO directly observes the $1.42~\mathrm{PeV}$ highest-energy photon. By adopting the synchrotion self-Compton model, LHAASO also suggests that the $1.12~\mathrm{PeV}$ high-energy photon from Crab Nebula corresponds to a $2.3~\mathrm{PeV}$ high-energy electron. We study the $1.42~\mathrm{PeV}$ photon decay and the $2.3~\mathrm{PeV}$ electron decay to perform a joint analysis on photon and electron two-dimensional Lorentz violation~(LV) parameter plane. Our analysis is systematic and comprehensive, and we naturally get the strictest constraints from merely considering photon LV effect in photon decay and electron LV effect in electron decay. Our result also permits the parameter space for new physics beyond relativity.
\end{abstract}



\begin{keyword}
ultrahigh-energy cosmic photon \sep Lorentz violation \sep photon decay \sep electron decay \sep 
photon and electron Lorentz violation parameter plane



\end{keyword}

\end{frontmatter}



The Large High Altitude Air Shower Observatory~(LHAASO) is a new-generation mountain observatory with unprecedented ultrahigh-energy~(UHE) photon detection capability~\cite{H1-cao-2021-ultrahigh,H8-cao-2021-pata}. Recently, LHAASO reported more than 530 UHE photons with energies larger than $100~\mathrm{TeV}$ from twelve astrophysical gamma-ray sources within the Milky Way, including the highest-energy photon detected at about $1.42~\mathrm{PeV}$~\cite{H1-cao-2021-ultrahigh}. LHAASO also reported the detection of gamma-ray spectrum from $5\times10^{-4}~\mathrm{PeV}$ to $1.12~\mathrm{PeV}$ from Crab Nebula, and these UHE photons exhibit the presence of a PeV electron accelerator~\cite{H8-cao-2021-pata}. These high-energy photons not only help to study the origin and acceleration mechanism of UHE cosmic rays, but also provide the opportunity to test fundamental physics concepts such as Lorentz symmetries of photons~\cite{H2-li-2021-ultrahigh,H3-lhaaso-2021-exploring} and electrons~\cite{H7-li-2022-testing}. By analysing the LHAASO $1.42~\mathrm{PeV}$ highest-energy photon~\cite{H1-cao-2021-ultrahigh}, Ref.~\cite{H2-li-2021-ultrahigh} got a photon superluminal linear Lorentz violation~{LV} constraint --- $E_\mathrm{LV}^\mathrm{(p,sup)}\ge2.74\times10^{33}~\mathrm{eV}$ from photon decay research. By analysing two gamma-ray sources LHAASO J2032+4102 and J0534+2202, LHAASO collaboration got a photon superluminal linear LV constraint $E_\mathrm{LV}^\mathrm{(p,sup)}\ge1.42\times10^{33}~\mathrm{eV}$ from photon decay research~\cite{H3-lhaaso-2021-exploring}. By analysing the Crab Nebula $1.12~\mathrm{PeV}$ highest-energy photon, Ref.~\cite{H7-li-2022-testing} obtained the most strict constraint on electron superluminal linear modified parameter $ E_\mathrm{LV}^\mathrm{(e,sup)}\ge9.4\times10^{25}~\mathrm{GeV}$ from the electron decay research. \\

In conventional case of relativity, there are no photon decay and electron decay phenomena in vacuum, but in LV case, things might be different. If photon or electron does decay, it is a sign for photon or electron LV effect. On the other hand, the UHE photon and electron data set very strict constraints on photon and electron LV effects. In previous photon decay analyses~\cite{H2-li-2021-ultrahigh,H3-lhaaso-2021-exploring}, only initial photon LV effect is considered, with LV effect for the out-going electron-positron pair neglected. But since the out-going particles obtain the whole energy-momentum of initial photon, it is necessary to consider the LV effect of out-going electron-positron pair. In previous electron decay analysis~\cite{H7-li-2022-testing}, there is a supposition that the emitted photon is soft enough that its LV effect can be neglected. But if we check the electron decay reaction we can find that: in some case, the out-going photon can obtain the half energy-momentum of the initial electron, so it is also necessary to consider the out-going photon LV effect in electron decay. \\

To get a systematic and comprehensive analysis on photon and electron LV parameter plane from LHAASO results, we restudy photon decay and electron decay. As the LV effects are very tiny, we introduce tiny LV modifications on the photon and electron dispersion relations:
\begin{equation}\label{dispersion relation of photon/electron by xi/eta}
    \begin{cases}
    \omega^2=k^2[1+\xi_\mathrm{n}(\frac{k}{E_\mathrm{Pl}})^n]   &  \mathrm{photon};\\
    E^2=m^2+p^2[1+\eta_\mathrm{n}(\frac{p}{E_\mathrm{Pl}})^n]  &  \mathrm{electron/positron}, \\
    \end{cases}
\end{equation}
where $E_\mathrm{Pl}$ is the Plank scale, and $\xi_\mathrm{n}$, $\eta_\mathrm{n}$ are the $n$th-order LV parameters of photon and electron respectively. For a threshold reaction, when the final particle momenta are parallel and the initial momenta are antiparallel, the threshold occurs~\cite{H6-Mattingly-2002-threshold}, so we only consider the modulus of the momentum $k=|\vec{k}|$, and $p=|\vec{p}|$, to obtain the threshold. If we only consider the linear modification~($n=1$), we simplify the notation: $\xi_\mathrm{1}:\equiv\xi$ and $\eta_\mathrm{1}:\equiv\eta$. \\

For photon decay $\gamma\to e^-+e^+$, considering a high-energy photon with momentum $k$ that decays into an electron with momentum $xk$~($x\in[0,1]$) and a positron with momentum $(1-x)k$, using photon and electron dispersion relations Eq.~(\ref{dispersion relation of photon/electron by xi/eta}) and the energy-momentum conservation relation, only considering the linear~($n=1$) modification, and expanding it to the leading-order of the LV parameters and the leading-order of $(m/k)^2$, we get~\cite{H12-jacobson-2003-threshold}:
\begin{equation}\label{energy-momentum conservation relation for photon decay II}
k[1+\frac{\xi}{2}\frac{k}{E_\mathrm{Pl}}]=xk[1+\frac{m^2}{2(xk)^2}+\frac{\eta}{2}\frac{xk}{E_\mathrm{Pl}}]+\{x\leftrightarrow 1-x\}.
\end{equation}
After simple algebraic operations, Eq.~(\ref{energy-momentum conservation relation for photon decay II}) becomes~\cite{H12-jacobson-2003-threshold}:
\begin{equation}\label{e-m relation of photon decay for linear modification}
    \frac{m^2E_\mathrm{Pl}}{k^3}=x(1-x)[\xi-((1-x)^2+x^2)\eta].
\end{equation}
Finding the threshold of the photon decay is equivalent to finding the minimum value of $k$ on the left side of the Eq.~(\ref{e-m relation of photon decay for linear modification}). When $\xi\to0$ and $\eta\to0$, $k\to+\infty$, and it is the classic case where photon cannot decay, and that is to be expected, since any case must go back to classical case when the LV effects do not at work. Then we get the thresholds in different LV parameter configurations~\cite{H12-jacobson-2003-threshold,H13-Jacobson-2001-TeV,H16-Konopka-2002-observational}:
\begin{equation*}
    k_\mathrm{th}^\mathrm{p}=
    \begin{cases}
    +\infty   &  (a)\quad\xi-\eta<0\quad \mathrm{and}  \quad 2\xi-\eta<0;\\
    (\frac{8m^2E_\mathrm{Pl}}{2\xi-\eta})^{1/3}  &  (b)\quad\xi>0\quad \mathrm{and}  \quad 2\xi-\eta>0;\\
    (\frac{-8\eta m^2E_\mathrm{Pl}}{(\xi-\eta)^2})^{1/3}  &   (c)\quad\eta<\xi<0.\\
    \end{cases}
\end{equation*}
In case (a), $k_\mathrm{th}^\mathrm{p}=+\infty$ means that there is no photon decay. In case (b), the momenta of the out-going particles are equally distributed, that is to say, the out-going electron/positron gains half momentum from photon, so it is necessary to consider the out-going electron/positron LV effect. In case (c), the momentum distribution is not equal, and it is also necessary to consider the out-going particle LV effects since they get the whole momentum from photon.\\

When the photon energy exceeds the threshold, the photon decay can occur and result a rapid drop in photon energy. On the other hand, observing a photon, whose energy is $E_\mathrm{p}$, means that $E_\mathrm{p}$ does not reach the decay threshold $k_\mathrm{th}^\mathrm{p}$: $k_\mathrm{th}^\mathrm{p}>E_\mathrm{p}$. In different LV parameter configurations, it is same as~\cite{H12-jacobson-2003-threshold}:
\begin{equation*}
    \begin{cases}
    \mathrm{no\quad extra\quad constraints}   &  (a)\quad\xi-\eta<0\quad \mathrm{and}  \quad 2\xi-\eta<0;\\
    0<2\xi-\eta<\frac{8m^2E_\mathrm{Pl}}{E_\mathrm{p}^3}  &  (b)\quad\xi>0\quad \mathrm{and}  \quad 2\xi-\eta>0;\\
    0<\xi-\eta<\sqrt{\frac{-8m^2E_\mathrm{Pl}\eta}{E_\mathrm{p}^3}}  &   (c)\quad\eta<\xi<0.\\
    \end{cases}
\end{equation*}\\

\begin{figure}[H]
    \centering
    \includegraphics[scale=0.7]{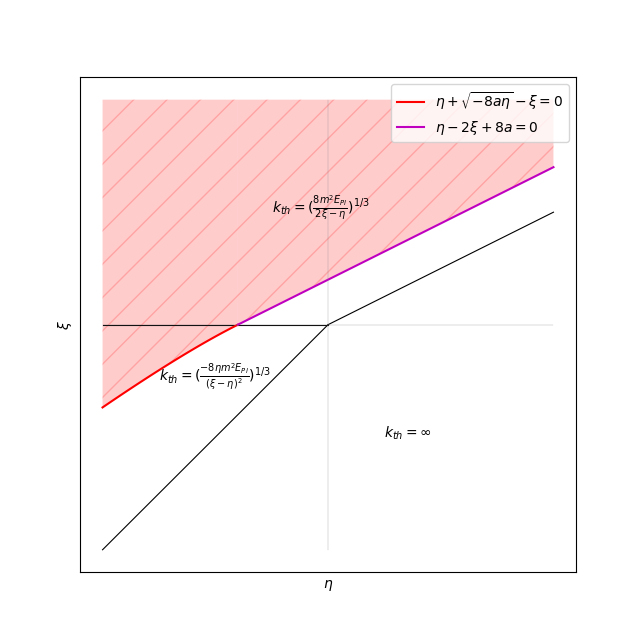}
    \caption{Photon decay constraint on photon-electron LV parameter plane from highest-energy photon~($a:\equiv m^2E_\mathrm{Pl}/E_\mathrm{p}^3$).}
    \label{fig:photon decay}
\end{figure}

From Fig.~\ref{fig:photon decay}, we clearly see that: 
\begin{itemize}
    \item In photon decay research, the highest-energy photon sets very strict constraint on $\xi>0$ range, while the $\xi<0$ range still has possible parameter space. The LHAASO $1.42~\mathrm{PeV}$ highest-energy photon sets very strict constraint on photon-electron LV parameter plane:
    \begin{equation*}
    \begin{cases}
    \mathrm{no\quad extra\quad constraints}   &  (a)\quad\xi-\eta<0\quad \mathrm{and}  \quad 2\xi-\eta<0;\\
    0<2\xi-\eta<8.90\times10^{-6}  &  (b)\quad\xi>0\quad \mathrm{and}  \quad 2\xi-\eta>0;\\
    0<\xi-\eta<\sqrt{-8.90\times10^{-6}\eta}  &   (c)\quad\eta<\xi<0.\\
    \end{cases}
    \end{equation*}
    \item If we presuppose that there is no electron LV effect~($\eta=0$), this assumption is what we usually use when we study photon decay, the constraint on photon LV parameter is $\xi<4m^2E_\mathrm{Pl}/E_\mathrm{p}^3$, and it is a strict constraint on photon superluminal linear LV modification. Considering the LHAASO $1.42~\mathrm{PeV}$ highest-energy photon, we get $\xi<4.45\times10^{-6}$. With the experimental errors considered, the constraint given by the $1.42^{+0.13}_{-0.13}~\mathrm{PeV}$ highest-energy photon is $\xi<4.45^{+1.49}_{-1.03}\times10^{-6}$.
    \item If we presuppose that there is no photon LV effect~($\xi=0$), the constraint on electron LV parameter is $\eta>-8m^2E_\mathrm{Pl}/E_\mathrm{p}^3$, and it is a strict constraint on electron subluminal linear modification. Considering the LHAASO $1.42~\mathrm{PeV}$ highest-energy photon, we get $\eta>-8.90\times10^{-6}$. With the experimental errors taken into account, the constraint given by the $1.42^{+0.13}_{-0.13}~\mathrm{PeV}$ highest-energy photon is $\eta>-8.90^{+2.06}_{-2.97}\times10^{-6}$.
\end{itemize}

Different from highest-energy photon, it is hard to ascertain the highest-energy electron from LHAASO data. The Crab Nebula UHE photon spectrum from LHAASO observation means that the Crab Nebula operates as an electron PeVatron~\cite{H8-cao-2021-pata}. According to the synchrotion self-Compton model~\cite{H9-atoyan-1996-mechanisms,H10-Meyer-2010-crab}\footnote{The synchrotion self-Compton model is a likely mechanism to produce the observed $1.12~\mathrm{PeV}$ photon, and there are also possibilities that the photon might be produced in some other mechanisms: for example, hadronic interaction of high-energy protons~(or nuclei) via neutral pion decay.}, the Crab Nebula gamma-ray above $\sim1~\mathrm{GeV}$ is produced via inverse-Compton process by UHE electron, so the Crab Nebula $1.12~\mathrm{PeV}$ highest-energy photon means the energy of the parent electron can naturally be considered as $>1~\mathrm{PeV}$~\cite{H7-li-2022-testing}. Through systematic analysis, LHAASO got a simple relation between the upscattered photon $E^\mathrm{\gamma}$ and the parent electron $E^\mathrm{e}$: $E^\mathrm{e}=2.15(E^\mathrm{\gamma}/1~\mathrm{PeV})^{0.77}~\mathrm{PeV}$~\cite{H8-cao-2021-pata}. Thus, for the Crab Nebula $1.12~\mathrm{PeV}$ highest-energy photon, the energy of the parent electron is $2.3~\mathrm{PeV}$~\cite{H8-cao-2021-pata}. Under the assumption of the synchrotion self-Compton model, $2.3~\mathrm{PeV}$ electron is the highest-energy electron that we can get from LHAASO data.\\

For electron decay $e^-\to e^-+\gamma$, considering a high-energy electron with momentum $k$ that decays into an electron with momentum $yk$~($y\in[0,1]$) and a photon with momentum $(1-y)k$, similarly to photon decay, we get:
\begin{equation}\label{energy-momentum conservation relation for electron decay II}
    k[1+\frac{m^2}{2k^2}+\frac{\eta}{2}\frac{k}{E_\mathrm{Pl}}]=yk[1+\frac{m^2}{2(yk)^2}+\frac{\eta}{2}\frac{yk}{E_\mathrm{Pl}}]+(1-y)k[1+\frac{\xi}{2}\frac{(1-y)k}{E_\mathrm{Pl}}],
\end{equation}
and~\cite{H12-jacobson-2003-threshold}:
\begin{equation}\label{e-m relation of electron decay for linear modification}
    \frac{m^2E_\mathrm{Pl}}{k^3}=y(y+1)\eta-y(1-y)\xi.
\end{equation}
When $\xi\to0$ and $\eta\to0$, $k\to+\infty$, and it is the classic case where electron cannot decay. We get the thresholds in different LV parameter configurations~\cite{H12-jacobson-2003-threshold,H13-Jacobson-2001-TeV,H16-Konopka-2002-observational}:
\begin{equation*}
    k_\mathrm{th}^\mathrm{e}=
    \begin{cases}
    +\infty   &  (i)\quad\eta<0\quad \mathrm{and}  \quad \xi-\eta>0;\\
    (\frac{m^2E_\mathrm{Pl}}{2\eta})^{1/3}  &  (j)\quad\eta>0\quad \mathrm{and}  \quad \xi+3\eta>0;\\
    (\frac{-4(\xi+\eta)m^2E_\mathrm{Pl}}{(\xi-\eta)^2})^{1/3}  &   (k)\quad\xi-\eta<0\quad \mathrm{and}  \quad \xi+3\eta<0.\\
    \end{cases}
\end{equation*}
In case (i), $k_\mathrm{th}^\mathrm{e}=+\infty$ means that there is no electron decay. In case (j), the out-going electron gains almost all of momentum. In case (k), the out-going photon gets a part of momentum from initial electron, so it is necessary to consider the photon LV effect in electron decay. \\

When the electron energy exceeds the threshold, the electron decay can occur and result a rapid drop in electron energy. On the other hand, observing an electron, whose energy is $E_\mathrm{e}$, means that $E_\mathrm{e}$ does not reach the decay threshold $k_\mathrm{th}^\mathrm{e}$: $k_\mathrm{th}^\mathrm{e}>E_\mathrm{e}$. In different LV parameter configurations, it is same as~\cite{H12-jacobson-2003-threshold}:
\begin{equation*}
    \begin{cases}
    \mathrm{no\quad extra\quad constraints}   &  (i)\quad\eta<0\quad \mathrm{and}  \quad \xi-\eta>0;\\
    \eta<\frac{m^2E_\mathrm{Pl}}{2E_\mathrm{e}^3}  &  (j)\quad\eta>0\quad \mathrm{and}  \quad \xi+3\eta>0;\\
    -\sqrt{-\frac{4m^2E_\mathrm{Pl}(\xi+\eta)}{E_\mathrm{e}^3}}<\xi-\eta<0  &   (k)\quad\xi-\eta<0\quad \mathrm{and}  \quad \xi+3\eta<0.\\
    \end{cases}
\end{equation*}

\begin{figure}[H]
    \centering
    \includegraphics[scale=0.7]{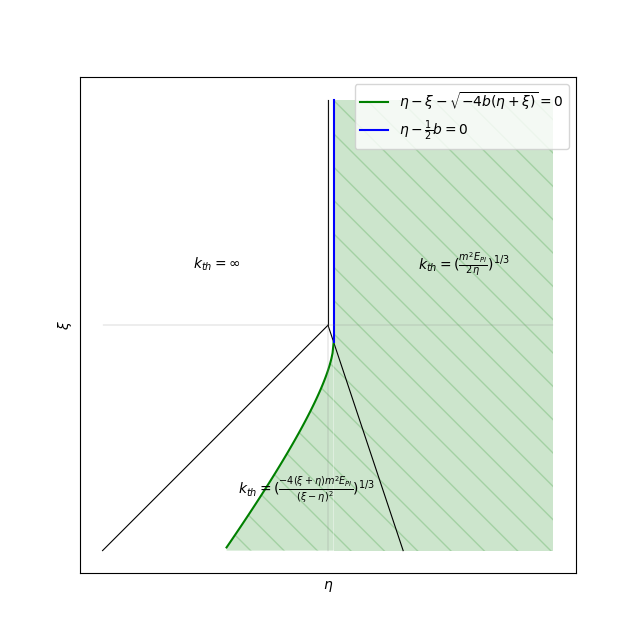}
    \caption{Electron decay constraint on photon-electron LV parameter plane from highest-energy electron~($b:\equiv m^2E_\mathrm{Pl}/E_\mathrm{e}^3$).}
    \label{fig:electron decay}
\end{figure}

From Fig. \ref{fig:electron decay}, we clearly see that: 
\begin{itemize}
    \item In electron decay research, the highest-energy electron sets very strict constraint on $\eta>0$ range, while the $\eta<0$ range still has possible parameter space. The LHAASO $2.3~\mathrm{PeV}$ highest-energy electron sets very strict constraint on photon-electron LV parameter plane:
    \begin{equation*}
    \begin{cases}
    \mathrm{no\quad extra\quad constraints}   &  (i)\quad\eta<0\quad \mathrm{and}  \quad \xi-\eta>0;\\
    \eta<1.3\times10^{-7}  &  (j)\quad\eta>0\quad \mathrm{and}  \quad \xi+3\eta>0;\\
    -\sqrt{-1.0\times10^{-6}(\xi+\eta)}<\xi-\eta<0  &   (k)\quad\xi-\eta<0\quad \mathrm{and}  \quad \xi+3\eta<0.\\
    \end{cases}
    \end{equation*} 
    \item If we presuppose that there is no electron LV effect~($\eta=0$), the constraint on photon LV parameter is $\xi>-4m^2E_\mathrm{Pl}/E_\mathrm{e}^3$, and it is a strict constraint on photon subluminal linear LV modification. Considering the LHAASO $2.3~\mathrm{PeV}$ highest-energy electron, we get $\xi>-1.0\times10^{-6}$.
    \item If we presuppose that there is no photon LV effect~($\xi=0$), this assumption is what we usually use when we study electron decay, the constraint on electron LV parameter is $\eta<m^2E_\mathrm{Pl}/(2E_\mathrm{e}^3)$, and it is a strict constraint on electron superluminal linear LV modification. Considering the LHAASO $2.3~\mathrm{PeV}$ highest-energy electron, we get $\eta<1.3\times10^{-7}$.
\end{itemize}

\begin{figure}[H]
    \centering
    \includegraphics[scale=0.7]{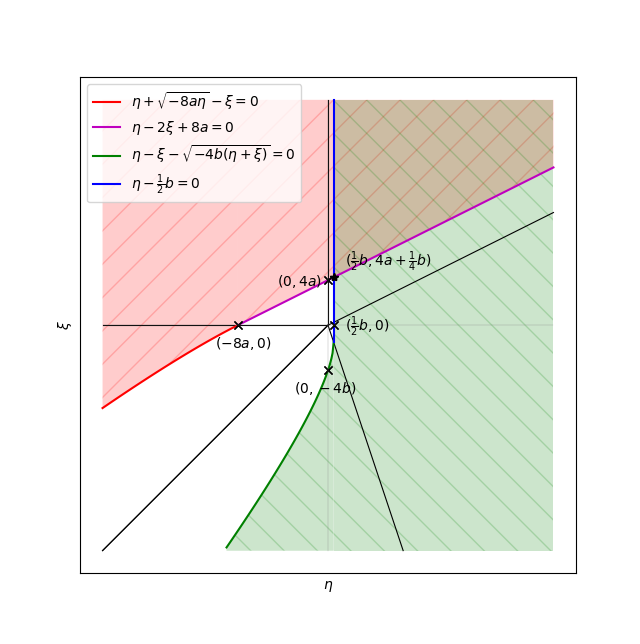}
    \caption{Joint constraint on photon-electron LV parameter plane from highest-energy photon and electron~($a:\equiv m^2E_\mathrm{Pl}/E_\mathrm{p}^3$ and $b:\equiv m^2E_\mathrm{Pl}/E_\mathrm{e}^3$).}
    \label{fig:joint limitation}
\end{figure}

Considering the photon decay and electron decay at the same time, we get:~(i) the highest-energy photon and electron set very strict constraints on 1st, 2nd and 4th quadrants of photon-electron plane. (ii) The limitation pole is $(\eta, \xi)<(m^2E_\mathrm{Pl}/(2E_\mathrm{e}^3), m^2E_\mathrm{Pl}/(4E_\mathrm{e}^3)+4m^2E_\mathrm{Pl}/E_\mathrm{p}^3)$. Considering the LHAASO $1.42~\mathrm{PeV}$ highest-energy photon and $2.3~\mathrm{PeV}$ highest-energy electron, we can get $(\eta, \xi)<(1.3\times10^{-7}, 4.5\times10^{-6})$. (iii) In the 3rd quadrant, the parameter space still has a lot of options. \\

We check the constraints on photon-electron LV parameter plane by LV scale parametric method:
\begin{equation}\label{dispersion relation of photon/electron by E_LV}
    \begin{cases}
    \omega^2=k^2(1-s^\mathrm{p}\frac{k}{E_\mathrm{LV}^\mathrm{p}})   &  \mathrm{photon};\\
    E^2=m^2+p^2(1-s^\mathrm{e}\frac{p}{E_\mathrm{LV}^\mathrm{e}})  &  \mathrm{electron/positron}, \\
    \end{cases}
\end{equation}
where $E_\mathrm{LV}^\mathrm{p}$ and $E_\mathrm{LV}^\mathrm{e}$ are the LV scale of photon and electron at which the LV effects become significant. $s^\mathrm{p/e}=+1$ means that the higher the energy the slower the speed of particles~(subluminal), and $s^\mathrm{p/e}=-1$ means faster~(superluminal). Using the correspondence between these two different parametric methods, we get previous strictest constraints on photon and electron LV parameters from the joint constraint on photon-electron LV parameter plane~(Fig. \ref{fig:joint limitation}):
\begin{itemize}
    \item There are very strict constraints on 1st, 2nd and 4th quadrants of photon-electron plane, and these strict constraints are same as the strict constraints on superluminal case~(one of $s^\mathrm{p/e}$ is $-1$). The limitation pole $(\eta, \xi)<(m^2E_\mathrm{Pl}/(2E_\mathrm{e}^3), m^2E_\mathrm{Pl}/(4E_\mathrm{e}^3)+4m^2E_\mathrm{Pl}/E_\mathrm{p}^3)$ is same as the limitation pole on superluminal LV scale $(E_\mathrm{LV}^\mathrm{e, sup}, E_\mathrm{LV}^\mathrm{p,sup})>(2E_\mathrm{e}^3/m^2, 4E_\mathrm{p}^3E_\mathrm{e}^3/(16m^2E_\mathrm{e}^3+m^2E_\mathrm{p}^3))$. Using the LHAASO $1.42~\mathrm{PeV}$ highest-energy photon and $2.3~\mathrm{PeV}$ highest-energy electron, we get $(E_\mathrm{LV}^\mathrm{e, sup}, E_\mathrm{LV}^\mathrm{p,sup})>(9.4\times10^{34}\mathrm{eV}, 2.7\times10^{33}\mathrm{eV})$. The 3rd quadrant is the subluminal case~($s^\mathrm{p}=s^\mathrm{e}=+1$), and this parameter space still has a lot of options.
    \item In case of no electron LV effect~($\eta=0$), the low limitation on photon LV parameter~($\xi>-4m^2E_\mathrm{Pl}/E_\mathrm{e}^3$) is from electron decay, and it is a photon subluminal LV scale limitation: $E_\mathrm{LV}^\mathrm{p, sub}>E_\mathrm{e}^3/(4m^2)$. Using the LHAASO $2.3~\mathrm{PeV}$ highest-energy electron, we get $E_\mathrm{LV}^\mathrm{p,sub}>1.2\times10^{34}~\mathrm{eV}$. The upper limitation~($\xi<4m^2E_\mathrm{Pl}/E_\mathrm{p}^3$) is from photon decay, and it is a photon superluminal LV scale limitation: $E_\mathrm{LV}^\mathrm{p,sup}>E_\mathrm{p}^3/(4m^2)$. Using the LHAASO $1.42~\mathrm{PeV}$ highest-energy photon, we get $E_\mathrm{LV}^\mathrm{p,sup}>2.74\times10^{33}~\mathrm{eV}$\footnote{With the experimental errors taken into account, the constraint given by the $1.42^{+0.13}_{-0.13}~\mathrm{PeV}$ highest-energy photon is $E_\mathrm{LV}^\mathrm{p,sup}>2.74^{+0.83}_{-0.68}\times10^{33}~\mathrm{eV}$.}, that is same as the strictest constraint from Ref.~\cite{H2-li-2021-ultrahigh}. This sameness is natural as these two parametric methods have no essential difference but in different expressions~\cite{H18-He-2022-lorentz}, and this strictest constraint is got under the supposition $\eta=0$.
    \item In case of no photon LV effect~($\xi=0$), the low limitation on electron LV parameter~($\eta>-8m^2E_\mathrm{Pl}/E_\mathrm{p}^3$) is from photon decay, and it is an electron subluminal LV scale limitation: $E_\mathrm{LV}^\mathrm{e,sub}>E_\mathrm{p}^3/(8m^2)$. Using the LHAASO $1.42~\mathrm{PeV}$ highest-energy photon, we get $E_\mathrm{LV}^\mathrm{e,sub}>1.37\times10^{33}~\mathrm{eV}$\footnote{With the experimental errors taken into account, the constraint given by the $1.42^{+0.13}_{-0.13}~\mathrm{PeV}$ highest-energy photon is $E_\mathrm{LV}^\mathrm{e,sub}>1.37^{+0.41}_{-0.34}\times10^{33}~\mathrm{eV}$.}. The upper limitation~($\eta<m^2E_\mathrm{Pl}/(2E_\mathrm{e}^3)$) is from electron decay, and it is an electron superluminal LV scale limitation: $E_\mathrm{LV}^\mathrm{e,sup}>2E_\mathrm{e}^3/m^2$. Using the LHAASO $2.3~\mathrm{PeV}$ highest-energy electron, we get $E_\mathrm{LV}^\mathrm{e,sup}>9.4\times10^{34}~\mathrm{eV}$, that is same as the strictest constraint from Ref.~\cite{H7-li-2022-testing}. This sameness is natural as these two parametric methods have no essential difference but in different expressions~\cite{H18-He-2022-lorentz}, and this strictest constraint is got under the supposition $\xi=0$.
\end{itemize}

Restudying the photon decay and electron decay, we know that it is necessary to consider both photon and electron LV effects. We perform a joint analysis on the photon-electron two-dimensional LV parameter plane. This analysis is based on the $1.42~\mathrm{PeV}$ highest-energy photon directly observed by LHAASO and the $2.3~\mathrm{PeV}$ highest-energy electron that corresponds to the Crab Nebula $1.12~\mathrm{PeV}$ photon observed by LHAASO through the synchrotion self-Compton model.
LHAASO $1.42~\mathrm{PeV}$ highest-energy photon and $2.3~\mathrm{PeV}$ highest-energy electron set very strict constraints on 1st, 2nd and 4th quadrants of photon-electron plane, whereas the 3rd quadrant still has permissible parameter space for new physics beyond relativity. This joint analysis of photon-electron LV parameter plane is systematic and comprehensive, and we naturally get the previous results of strictest constraints on photon and electron LV effects.\\

{\bf{Acknowledgement}} This work is supported by National Natural Science Foundation of China (Grant No.~12075003).






\end{document}